# A Benchmark JWST Near-Infrared Spectrum for the Exoplanet WASP-39b

A. L. Carter[1,2]* & E. M. May[3]*, N. Espinoza[1,4], L. Welbanks[5,6], E. Ahrer[7,8,9], L. Alderson[10], R. Brahm[11,12,13], A. D. Feinstein[14,15], D. Grant[10], M. Line[5], G. Morello[16,17], R. O'Steen[1], M. Radica[18,19], Z. Rustamkulov[20], K. B. Stevenson[3], J. D. Turner[21,6], M. K. Alam[1,22,], D. R. Anderson[8,7], N. M. Batalha[2], M. P. Battley[23], D. Bayliss[8], J. L. Bean[14], B. Benneke[24], Z. K. Berta-Thompson[25], J. Brande[26], E. M. Bryant[27], M. R. Burleigh[28], L. Coulombe[24], I. J.M. Crossfield[26], M. Damiano[29], J.-M. Désert[30], L. Flagg[21], S. Gill[8,7], J. Inglis[31], J. Kirk[32,33], H. Knutson[31], L. Kreidberg[9], M. López Morales[34], M. Mansfield[35,6], S. E. Moran[36], C. A. Murray[25], M. C. Nixon[37], D. J.M. Petit dit de la Roche[23], B. V. Rackham[38,39,40], E. Schlawin[35], D. K. Sing[20,41], H. R. Wakeford[10], N. L. Wallack[22], P. J. Wheatley[7,8], S. Zieba[9,42], K. Aggarwal[43], J. K. Barstow[44], T. J. Bell[45], J. Blecic[46,47], C. Caceres[48,49,50], N. Crouzet[42], P. E. Cubillos[51,52], T. Daylan[53,54], M. de Val-Borro[55], L. Decin[56], J. J. Fortney[2], N. P. Gibson[57], K. Heng[58,8], R. Hu[29,31], E M.-R. Kempton[37], P. Lagage[59], J. D. Lothringer[1,60], J. Lustig-Yaeger[3], L. Mancini[61,62,9], N. J. Mayne[63], L. C. Mayorga[3], K. Molaverdikhani[58,64], E. Nasedkin[9], K. Ohno[65], V. Parmentier[66], D. Powell[34,6], S. Redfield[67], P. Roy[24], J. Taylor[68,24], X. Zhang[69]

[1]Space Telescope Science Institute, 3700 San Martin Drive, Baltimore, MD 21218, USA
[2]Department of Astronomy & Astrophysics, University of California, Santa Cruz, Santa Cruz, CA, USA
[3]Johns Hopkins APL, Laurel, MD, USA
[4]Department of Physics & Astronomy, Johns Hopkins University, Baltimore, MD 21218, USA
[5]School of Earth and Space Exploration, Arizona State University, Tempe, AZ, USA
[6]NHFP Sagan Fellow
[7]Centre for Exoplanets and Habitability, University of Warwick, Coventry, UK
[8]Department of Physics, University of Warwick, Coventry, UK
[9]Max Planck Institute for Astronomy, Heidelberg, Germany
[10]School of Physics, University of Bristol, Bristol, UK
[11]Facultad de Ingeniería y Ciencias, Universidad Adolfo Ibáñez, Av. Diagonal las Torres 2640, Peñalolén, Santiago, Chile
[12]Millennium Institute for Astrophysics, Chile
[13]Data Observatory Foundation, Chile
[14]Department of Astronomy & Astrophysics, University of Chicago, Chicago, IL, USA
[15]Laboratory for Atmospheric and Space Physics, University of Colorado Boulder, Boulder, CO, USA
[16]Department of Space, Earth and Environment, Chalmers University of Technology, Gothenburg, Sweden
[17]Instituto de Astrofísica de Canarias (IAC), Tenerife, Spain
[18]Department of Physics, Université de Montréal, Montréal, Québec, Canada
[19]Trottier Institute for Research on Exoplanets, Université de Montréal, Montréal, Québec, Canada
[20]Department of Earth and Planetary Sciences, Johns Hopkins University, Baltimore, MD, USA
[21]Department of Astronomy and Carl Sagan Institute, Cornell University, Ithaca, NY, USA
[22]Earth and Planets Laboratory, Carnegie Institution for Science, Washington, DC, USA
[23]Département d'Astronomie, Université de Genève, Sauverny, Switzerland
[24]Department of Physics and Institute for Research on Exoplanets, Université de Montréal, Montreal, QC, Canada
[25]Department of Astrophysical and Planetary Sciences, University of Colorado, Boulder, CO, USA
[26]Department of Physics & Astronomy, University of Kansas, Lawrence, KS, USA
[27]Mullard Space Science Laboratory, University College London, Holmbury St Mary, Dorking, Surrey, UK
[28]School of Physics and Astronomy, University of Leicester, Leicester

[29]Astrophysics Section, Jet Propulsion Laboratory, California Institute of Technology, Pasadena, CA, USA
[30]Anton Pannekoek Institute for Astronomy, University of Amsterdam, Amsterdam, The Netherlands
[31]Division of Geological and Planetary Sciences, California Institute of Technology, Pasadena, CA, USA
[32]Department of Physics, Imperial College London, London, UK
[33]Imperial College Research Fellow
[34]Center for Astrophysics | Harvard & Smithsonian, Cambridge, MA, USA
[35]Steward Observatory, University of Arizona, Tucson, AZ, USA
[36]Lunar and Planetary Laboratory, University of Arizona, Tucson, AZ, USA.
[37]Department of Astronomy, University of Maryland, College Park, MD, USA
[38]Department of Earth, Atmospheric and Planetary Sciences, Massachusetts Institute of Technology, Cambridge, MA, USA
[39]Kavli Institute for Astrophysics and Space Research, Massachusetts Institute of Technology, Cambridge, MA, USA
[40]51 Pegasi b Fellow
[41]Department of Physics & Astronomy, Johns Hopkins University, Baltimore, MD, USA
[42]Leiden Observatory, University of Leiden, Leiden, The Netherlands
[43]Indian Institute of Technology, Indore, India
[44]School of Physical Sciences, The Open University, Milton Keynes, UK
[45]BAER Institute, NASA Ames Research Center, Moffet Field, CA, USA
[46]Department of Physics, New York University Abu Dhabi, Abu Dhabi, UAE
[47]Center for Astrophysics and Space Science (CASS), New York University Abu Dhabi, Abu Dhabi, UAE
[48]Instituto de Astrofisica, Universidad Andres Bello, Santiago, Chile
[49]Centro de Astrofisica y Tecnologias Afines (CATA), Casilla 36-D, Santiago, Chile
[50]Nucleo Milenio de Formacion Planetaria (NPF), Chile
[51]INAF – Osservatorio Astrofisico di Torino, Pino Torinese, Italy
[52]Space Research Institute, Austrian Academy of Sciences, Graz, Austria
[53]Department of Astrophysical Sciences, Princeton University, Princeton, NJ, USA
[54]LSSTC Catalyst Fellow
[55]Planetary Science Institute, Tucson, AZ, USA
[56]Institute of Astronomy, Department of Physics and Astronomy, KU Leuven, Leuven, Belgium
[57]School of Physics, Trinity College Dublin, Dublin, Ireland
[58]Universitäts-Sternwarte, Ludwig-Maximilians-Universität München, München, Germany
[59]Université Paris-Saclay, Université Paris Cité, CEA, CNRS, AIM, Gif-sur-Yvette, France
[60]Department of Physics, Utah Valley University, Orem, UT, USA
[61]Department of Physics, University of Rome "Tor Vergata", Rome, Italy
[62]INAF - Turin Astrophysical Observatory, Pino Torinese, Italy
[63]Department of Physics and Astronomy, University of Exeter, Exeter, Devon, United Kingdom.
[64]Exzellenzcluster Origins, Garching, Germany
[65]Division of Science, National Astronomical Observatory of Japan, Mitaka-shi, Tokyo, Japan
[66]Université Côte d'Azur, Observatoire de la Côte d'Azur, CNRS, Laboratoire Lagrange, France
[67]Astronomy Department and Van Vleck Observatory, Wesleyan University, Middletown, CT, USA
[68]Atmospheric, Oceanic and Planetary Physics, Department of Physics, University of Oxford, Oxford, UK
[69]Department of Earth and Planetary Sciences, University of California Santa Cruz, Santa Cruz, California, USA

**Observing exoplanets through transmission spectroscopy supplies detailed information on their atmospheric composition, physics, and chemistry. Prior to *JWST*, these observations were limited to a narrow wavelength range across the near-ultraviolet to near-infrared, alongside broadband photometry at longer wavelengths. To understand more complex properties of exoplanet atmospheres, improved wavelength coverage and resolution are necessary to robustly quantify the influence of a broader range of absorbing molecular species. Here we present a combined analysis of *JWST* transmission spectroscopy across four different instrumental modes spanning 0.5–5.2 micron using**

**Early Release Science observations of the Saturn-mass exoplanet WASP-39b. Our uniform analysis constrains the orbital and stellar parameters within sub-percent precision, including matching the precision obtained by the most precise asteroseismology measurements of stellar density to-date, and further confirms the presence of Na, K, H$_2$O, CO, CO$_2$, and SO$_2$ atmospheric absorbers. Through this process, we also improve the agreement between the transmission spectra of all modes, except for the NIRSpec PRISM, which is affected by partial saturation of the detector. This work provides strong evidence that uniform light curve analysis is an important aspect to ensuring reliability when comparing the high-precision transmission spectra provided by JWST.**

WASP-39b has a mass of ~0.28 $M_{Jup}$, an equilibrium temperature of ~1100 K, and a highly inflated radius of ~1.27 $R_{Jup}$[14], making it an ideal target for transmission spectroscopy observations. Past optical and near-infrared observations of WASP-39b with ground-based telescopes, the *Hubble Space Telescope* (*HST*), and *Spitzer* have demonstrated evidence for strong absorption features that are not severely affected by the muting effects of cloud extinction[15-18], which has been mirrored in the initial data releases for each of our *JWST* observations across the near-infrared[5,7,10-12]. Furthermore, the late G-type host star WASP-39 (0.93±0.03 $M_\odot$, 0.895±0.23 $R_\odot$) is known to be relatively inactive, limiting the potential impact of stellar contamination[7,14,19,20]. The atmosphere of WASP-39b has a metal enrichment (metallicity) greater than that of its host star, although the range of metallicities which satisfy the observed *JWST* spectra differ between instrumental modes and can extend up to 100× solar metallicity[5,7,11,12]. Similarly, the measured ratio of carbon- and oxygen-bearing molecular species (C/O) in the atmosphere of WASP-39b seems to be sub-stellar or stellar, depending on the instrumental mode. These differing compositional measurements are likely due to the preliminary modeling performed and the sensitivities of specific wavelength regions to relevant molecular tracers. Measurements of the metallicity and C/O ratio for an exoplanet are important indicators of its bulk atmospheric chemistry[21-23] and formation history[24-26]; therefore, these *JWST* observations must be analyzed in a homogeneous manner so that the complementary constraining power of their different resolutions and wavelength ranges can be fully realized, and the nature of WASP-39b can be best understood.

We begin our data analysis with the extracted spectral time series as reported in the initial data release publications for these observations[5,7,11,12]. At this initial stage, the data have been corrected for both background and $1/f$ noise (additional correlated read noise due to, for example, biases in the detector readout electronics) when necessary. As there are multiple comparable reductions presented in each of these publications, we select the nominal case from each. Specifically, we choose the `supreme-SPOON`[5,27] reduction for Near-Infrared Imager and Slitless Spectrograph (NIRISS) Single Object Slitless Spectroscopy (SOSS)[3,4], the `Eureka!`[28] reduction for Near-Infrared Camera (NIRCam) F210M+F322W2[6,7], the `ExoTIC-JEDI` [V2][29] reduction for Near-Infrared Spectrograph (NIRSpec) G395H[8,9,12,13], and the `FIREFLy`[30] reduction for NIRSpec PRISM[8-11]. As WASP-39b is the only exoplanet to date to have been observed with such a diversity in instrumental capability, this is the first opportunity for a one-to-one cross-comparison between these modes and a verification of their relative performance. We extract seven separate white light curves from these data for our analysis using `Eureka!`[28]: two from the separate NIRISS SOSS Order 1 and Order 2 spectroscopy, one from the NIRCam F210M photometry, one from the NIRCam F322W2 spectroscopy, two from the NIRSpec G395H spectroscopy captured separately on the NRS1 and NRS2 detectors, and one from the NIRSpec PRISM spectroscopy (see Methods, Extended Figs 1-3, Extended Table 1).

To constrain the parameters for WASP-39b and its host star we perform a joint fit using `juliet`[31] to the seven *JWST* white light curves, in addition to: the "*Transiting Exoplanet Survey Satellite*" *(TESS)*[32] light curve, six separate "*Next Generation Transit Survey*" *(NGTS)*[33] light curves[7], and the CORALIE and "Spectrograph for the Observation of the PHenomena of stellar Interiors and Exoplanets" (SOPHIE) radial velocity measurements[14]. All light curve fits are displayed in Figure 1 and best-fit parameters are provided in the Table 1. With the combined constraining power of these data, we are able to obtain exquisite constraints on the WASP-39 system. The period of WASP-39b is constrained at sub-second precision (~0.3 seconds), with other physical and orbital parameters constrained at sub-percent precision (~0.1–0.5%). Of particular interest is the constraint on stellar density provided by the fitting, as it is constrained to ~0.3%—an equivalent precision to the most precise asteroseismology measurements made to date[34]. This is a direct consequence of the sampling of the transit events by the different *JWST* observations, which constrain the period and transit duration at unprecedented precision, and together with Kepler's third law, define the stellar density[35]. If such precisions are common for JWST white-light curves in general, then they could give rise to better constraints on orbital decay or transit timing variations, and improved stellar density measurements may improve constraints on system ages.

Spectrophotometric light curves are extracted at the native spectral resolution from their corresponding spectral time series using the `Eureka!`[28] package, with orbital parameters fixed to the best-fit values from the white light curve fitting. This results in 1028 light curves for NIRISS SOSS ($R\sim350\text{-}1390$, $\sigma_{mean}$=310 ppm), 550 for NIRCam F322W2 ($R\sim850\text{-}1360$, $\sigma_{mean}$=294 ppm), 1163 for NIRSpec G395H ($R\sim1340\text{-}2630$, $\sigma_{mean}$=496 ppm), 147 for NIRSpec PRISM ($R\sim20\text{-}290$, $\sigma_{mean}$=108 ppm), and an overall total of 2888 individual spectrophotometric light curves. As part of our analysis, we also investigate reductions at lower resolution binning schemes and find that unless the underlying limb-darkening parameters are fixed during light curve fitting, significant wavelength-dependent variations in excess of 150 ppm are present between native spectral resolution and $R$=100 spectra (see Methods, Extended Figure 4). The combined transmission spectra of the *JWST* observations from this work, alongside those of the initial data release publications, is displayed in Figure 2. The initial releases have different uncertainties at some locations as they are at a different resolution than the synthesized release.

The measured transmission spectra from the initial data releases display clear offsets relative to each other, whereas such offsets are reduced for the synthesized spectra after following the joint light curve fitting procedure described above (Figure 2, also see Methods). In the most extreme case, the NIRSpec G395H and NIRISS SOSS spectra have a mean offset in their overlapping region of 343±16 ppm when using the initial spectra, and 138±16 ppm when using the synthesized spectra. This indicates that different assumptions and inferences during the light curve fitting process can significantly affect the final measured transmission spectrum, even with the constraining power of a single, highly precise, *JWST* white light curve. Despite these synthesizing efforts, offsets are still present between the different modes. Median offsets of the spectra from the higher resolution modes relative to the NIRSpec PRISM spectrum are shown in Figure 2c. Differences in the median offsets between the higher resolution modes and NIRSpec PRISM are driven by the wavelength-dependent nature of NIRSpec PRISM systematics, compounded with wavelength-independent offsets between the higher resolution modes (see Methods, Extended Figure 5, Extended Table 2).

Wavelength-independent offsets are commonly seen between different telescopes and/or instruments and are typically due to different orbital or stellar parameters assumed, or different instrument sensitivities. However, an offset is present between NIRSpec PRISM and NIRSpec G395H, even though we use consistent orbital and stellar parameters, and the NIRSpec PRISM detector is the same as the short wavelength (NRS1) NIRSpec G395H detector. One potential explanation is that the low number of groups used in the NIRSpec PRISM observations may increase the influence of first-group effects and drive a more significant offset compared to the other observations, which utilize more groups. Such a shift has been observed in NIRCam transit observations[36] and may likely be present in this data also (see Methods). Upon applying a shift to the NIRSpec PRISM data it is possible to better match the other datasets across a broader wavelength range, however, a wavelength-dependent offset is still apparent from ~0.6–2.0 μm. It is across this wavelength range that the NIRSpec PRISM data are affected by detector saturation.

Despite an initial assessment that saturated data could be recovered for NIRSpec PRISM[11], there is a clear discrepancy when compared to the unsaturated NIRISS SOSS data. In reality, the complex interaction of detector saturation, non-linearity, first-group effects, and pixel cross-talk modulate the measured transit depth as a function of wavelength. An investigation into determining a potential further correction to the saturated data is presented in the Methods section, Extended Figures 6-8; however, for the purposes of our model analysis we rely solely on the NIRISS SOSS measurements at the wavelengths impacted by saturation. For future analyses, we recommend taking significant caution when inferring atmospheric properties from data that have been directly influenced by saturation. Such a conclusion is in agreement with past observations using earlier generations of infrared detectors, for which modeling the detector response to saturation was difficult[37], and data experiencing saturation were discarded[38]. Even if absorption features are present across a region of partial saturation that qualitatively match the predictions of an atmospheric model, the precise structure and amplitudes of those features may not be reliable. Where reliable inferences on the structure of an absorption feature are required, instead of obtaining saturated NIRSpec PRISM data, observers could consider using multiple higher-resolution modes. For example, NIRISS SOSS and NIRSpec G395M/H cover a similar wavelength range as NIRSpec PRISM but saturate more slowly due to their higher resolving powers. Alternatively, the NIRSpec G140M/H mode could be utilized alongside NIRSpec PRISM, as its ~0.97–1.89 μm wavelength range spans the region of NIRSpec PRISM that has the highest throughput and is most prone to saturation, but offers a ~2 magnitude improvement in brightness limit.

The combined transmission spectra exhibit a variety of spectroscopic features that can be attributed to absorption from elemental and molecular species. We investigate the origin of these spectroscopic features by comparing the observations to a self-consistent 1-dimensional radiative-convective photo-chemical-equilibrium (1D-RCPE) model that assumes a 10×solar metallicity and a sub-solar carbon-to-oxygen ratio of 0.35 (Figure 3); consistent with the inferred atmospheric properties from the initial data releases. We consider the possibility of inhomogeneous aerosols shaping the absorption features in our spectrum by post-processing the 1D-RCPE model with clouds and hazes resulting from a fit to the data (see Methods). Additionally, we allow for a uniform offset to the NIRSpec PRISM data relative to the chosen model and find a median value of -177 ppm. The resulting atmospheric model confirms that the spectral features are best explained by absorption due to Na, K, $H_2O$, $SO_2$, and $CO_2$. We also note the presence of a narrow absorption excess at ~1.083 μm that is evident across all explored binning schemes and may be indicative of absorption from metastable Helium[39,40], but a further investigation is outside the scope of this paper. The spectral fit additionally confirms the need for cloud extinction as expected from the relatively muted spectral features in the data. A more detailed analysis of these data, which: covers a broader variety of atmospheric modeling methodologies, explores offsets for all of the *JWST* modes, and provides constraints on accessible atmospheric properties, is presented in a companion publication (Welbanks et al. Submitted).

Importantly, we find that a uniform analysis, namely joint white light curve fitting for consistent orbital parameters, results in an improved agreement over previously independent analyses between all *JWST* observing modes considered, with the exception of NIRSpec PRISM. Caution should be exercised when combining *JWST* spectra from different instruments without a uniform light curve analysis, particularly if those datasets have been analyzed by independent teams. Until a more comprehensive limb-darkening investigation is performed, fixing limb-darkening parameters to models rather than fitting for these parameters at high resolution is necessary to improve consistency between spectra of the same dataset at different resolutions. Finally, while the included NIRSpec PRISM observations were impacted by several detector effects, particularly due to saturation (see Methods), this mode remains a powerful tool for efficient characterization of planets around dimmer stars.

The continued effort to understand how to best combine data from multiple instruments is important to accurately characterize exoplanet atmospheres. By combining their differing capabilities these broad-wavelength, high-precision, and high-resolution measurements will facilitate a wide range of model analyses, beginning with those presented in our companion publication (Welbanks et al. Submitted), and will greatly improve our understanding of the origins, histories, and atmospheres of exoplanets. Finally, with future ultraviolet and mid-infrared transmission measurements of WASP-39b also on the horizon (*HST* GO-17162, *JWST* DDT-2783), we are poised to begin exploring the full potential of this new era of exoplanet characterization and the scientific advances that it can offer.

## Methods

**Data Reduction**

The data presented in this work were obtained from a selection of observations of WASP-39b from the Panchromatic Transmission sub-program within The *JWST* Transiting Exoplanet Community Director's Discretionary ERS program[1,2] (ERS 1366; PIs: N. M. Batalha, J. L. Bean, K. B. Stevenson). This includes primary transit observations with: Near-Infrared Imager and Slitless Spectrograph (NIRISS) Single Object Slitless Spectroscopy (SOSS)[3-5] from July 26-27, 2022 (20:53 – 05:35 UT), Near-Infrared Camera (NIRCam) F210M+F322W2[6,7] from 22-23 July 2022 (19:28 – 03:40 UT), Near-Infrared Spectrograph (NIRSpec) G395H[8,9,12,13] from 30-31 July 2022 (21:45 – 06:21 UT), and NIRSpec PRISM[8-11] on 10 July, 2022 (15:05 – 23:39 UT). These observational modes span all three of *JWST*'s near-infrared instruments with resolving powers of $R \simeq 100$–2700 depending on the mode, and have overlapping wavelength ranges within a combined range of 0.518–5.348 μm.

The data reduction for this work begins with the extracted spectral time series as presented in the initial ERS publications for these observations[5,7,11,12]. At this stage, the data have undergone data processing steps such as detector-level corrections, ramp fitting, flat fielding, subtraction of background and $1/f$ noise, wavelength calibration, and spectral extraction. For a detailed account of the precise analysis steps taken for each instrumental mode, we refer the reader to the initial ERS publications.

As the initial ERS publications provide a variety of different reductions to the data, spanning different pipelines and different methodologies, we select just a single reduction from each for our analyses. For NIRCam F322W2, NIRSpec PRISM, and NIRSpec G395H, we select the reduction which matches that chosen in the initial publication, but for NIRISS SOSS we adopt a different reduction due to improvements in the out-of-transit baseline scatter. Specifically, this corresponds to the `Eureka!`[28] reduction for NIRCam F322W2, the `supreme-SPOON`[5,27] reduction for NIRISS SOSS, the `ExoTIC-JEDI [V2]`[29] reduction for NIRSpec G395H, and the `FIREFLy`[30] reduction for NIRSpec PRISM. We note that while a single reduction pipeline may be desirable, fundamentally different reduction procedures are required between instrumental modes, and a "jack of all trades" pipeline will not necessarily produce the most consistent results. Further, analyses from the initial data releases[5,7,11,12] demonstrated that different pipelines can reach a good agreement on the resulting spectra. Median out-of-transit stellar spectra for each of the selected reductions are displayed in Extended Data Figure 1. For the NIRCam F210M photometry, we do not repeat any data reduction procedures and adopt the existing extracted light curve[7]. Extended Data Table 1 gives an overview of the *JWST* observations included in this work.

**White Light Curve Analysis**

Seven separate white light curves were obtained from these *JWST* datasets: one from NIRSpec PRISM, two from NIRSpec G395H (one from each detector), two from NIRISS SOSS (one for each order), one from NIRCam F322W2, and one from the NIRCam F210M photometry. For the spectroscopic observations, white light curves are constructed using Eureka![28] across similar wavelength ranges to those adopted in the initial ERS publications[5,7,11,12], except for NIRSpec PRISM, where we exclude wavelengths below 2 μm due to the presence of saturation. This corresponds to wavelength ranges of: 0.873–2.808 μm for NIRISS SOSS Order 1, 0.6–0.9 μm for NIRISS SOSS Order 2, 2.420–4.025 μm for NIRCam F322W2, 2.725–3.716 μm for NIRSpec G395H NRS1, 3.829–5.172 μm for NIRSpec G395H NRS2, and 2.0–5.5 μm for NIRSpec PRISM.

We performed a joint fit to these light curves, in conjunction with six *NGTS* light curves, one *TESS* light curve (with 3 transits), and radial velocity measurements from CORALIE and SOPHIE[14]; these auxiliary datasets were selected as they were readily available, didn't show strong systematic effects and provided time-stamps we were able to join together with our *JWST* measurements. Radial velocities are mean-subtracted before performing any fitting. Dilution factors for each light curve are fixed to 1, implying we assume no dilution from nearby contaminants on those light curves. High-contrast imaging observations of WASP-39b reveal no nearby companions [41,42], and no contaminating sources are apparent in the NIRCam target acquisition image. In addition, the nominal joint fit presented and used in this work had eccentricity fixed to 0 (a fit leaving the eccentricity as a free parameter with the priors described below constrains it at $e < 0.039$ with 99% credibility; the rest of the posterior parameters being consistent at 1-sigma with the ones here presented). All time-stamps are converted to BJD TDB.

The free parameters in this fit included:

- The period, which had a normal prior distribution centered at the value reported in Maciejewski et al. 2016[43], i.e., 4.0552765 days but with a significantly larger standard deviation of 1 minute to allow for possible time-stamp mismatches between different BJD standards in the literature (e.g., UTC or TDB[44]).

- The time of transit-center, which had a normal prior centered at 2459791.615201 BJD TDB (the time of the NIRSpec/G395H observations) with a relatively large standard deviation of 0.1 days.

- The impact parameter, also centered at the value reported in Maciejewski et al. 2016[43], i.e., 0.45, but with a truncated normal distribution between 0 and 1, and with a larger standard deviation of 0.1.

- The stellar density, whose prior was set to a log-uniform distribution between 0.1 and 10 g/cm$^3$.

- The radial velocity semi-amplitude, which had a uniform prior between 0 and 200 m/s.

- An individual radial velocity offset for each radial velocity instrument with data in Faedi et al. 2011 (CORALIE and SOPHIE), with uniform priors between -100 and 100 m/s as well as jitter terms with log-uniform priors between 1 and 100 m/s for each.

- An individual planet-to-star radius ratio for each light curve, which had a uniform prior between 0 and 0.3, in order to account for possible wavelength-dependent planet-to-star radius ratio changes.

- The limb-darkening coefficients using a transformed quadratic law via the uninformative sampling prescription of Kipping et al. 2013 — which implied two parameters per light curve, $q_1$ and $q_2$, with uniform distributions between 0 and 1.

- A flux normalization term for each light curve, set with a normal prior centered at 0 and with a standard deviation of 100,000 ppm.

- A jitter term per light curve, set to a log-uniform prior between 0.1 and 10,000 ppm.

To handle instrumental systematics in the light curves, based on analyses performed on the out-of-transit data, we decided to use the following models:

- A Gaussian Process (GP) on the NIRCam/F322W2, NIRCam F210M photometry, NIRISS/SOSS Order 1 & 2 and *TESS* data. We chose a Matèrn 3/2 kernel and used time as the only regressor. The prior on the amplitude of this GP was set with a log-uniform distribution from 0.01 to 100 ppm for the *JWST* light curves and from 0.001 to 100 ppm for the *TESS* light curves. The timescale also had a log-uniform prior distribution between 0.01 and 100 days. The Bayesian evidence suggests that adding a GP to the NIRspec datasets does not provide an improvement to the fits.

- A linear model for the NIRSpec/G395H data with two regressors: a simple slope in time, and a regressor that was 0 before the tilt event observed in the data[12] and 1 after it.

- A linear model for NIRSpec/PRISM data with a simple slope in time.

Adding similar systematic models for the *NGTS* data didn't change the results of our fit. In total, 84 free parameters were used to fit a total of 12,206 data points, for which we used the dynamic nested sampling scheme as implemented in `dynesty`[45]. Some of the resulting parameters from this joint fit are presented in Table 1. Posteriors for select parameters are shown in Extended Data Figure 2.

**Wavelength Binning Scheme for Spectrophotometric Light Curve Extraction**

To extract the spectrophotometric light curves it is necessary to define a wavelength binning scheme for each dataset. The largest number of spectrophotometric bins, and therefore the highest resolution, is reached by binning at the native pixel resolution, where a spectrophotometric light curve is extracted for each individual pixel column. However, for these *JWST* modes, the native pixel resolution is higher than the native *spectral* resolution, which defines the difference in wavelengths, $\Delta\lambda$, that can be resolved at a given wavelength, $\lambda$. As such, we adopt the native spectral resolution as a fundamental baseline for the extraction of the spectrophotometric channels. While higher resolution schemes, including native pixel resolution, may theoretically offer access to narrower spectral features, understanding the potential and reliability of such an approach is beyond the scope of this work.

The native spectral resolving power, which defines the native spectral resolution, can be determined in units of pixels for each mode following $R_{\mathrm{pix}}=\lambda/DR$, where $\lambda$ is the wavelength, $D$ is the dispersion of the instrumental mode, and $R$ is the spectral resolving power of the instrumental mode. For all modes, we take the dispersion and resolving power curves from the reference data files provided by the *JWST* exposure time calculator, Pandeia[46]. Importantly, $R_{\mathrm{pix}}$ is a continuous function of wavelength, whereas individual pixel columns have discrete edges and cannot be meaningfully subdivided in wavelength. Therefore, we convert $R_{\mathrm{pix}}$ to integer pixel values using a ceiling function to ensure that pixel columns are not split across two separate wavelength bins. We define the bin edges beginning with the lowest wavelength pixel column, where the wavelength at the lower edge of this column corresponds to the lower edge of the first spectrophotometric bin. Pixel columns are then added to this bin until the number of columns is equal to $R_{\mathrm{pix}}$, and the wavelength at the upper edge of the final column corresponds to the upper edge of the first spectrophotometric bin. This process is repeated, using the previously determined upper edge as the starting lower edge for the next bin, until all bin edges have been defined. In the event that there is a transition in the integer value of $R_{\mathrm{pix}}$ as pixel columns are added to a bin, the highest value of $R_{\mathrm{pix}}$ is used to define the number of pixel columns that must be included. If there are not enough pixels available in the uppermost wavelength bin to satisfy this requirement, then those pixels are instead incorporated into the penultimate bin.

For each instrumental mode, we first extract spectrophotometric light curves at the native spectral resolution following this binning scheme across a subset of the full wavelength range using the values adopted in the initial ERS publications for each of the instrumental modes[5,7,11,12]. This corresponds to ranges of: 0.873–2.808 μm for NIRISS SOSS Order 1, 0.630–0.853 μm for NIRISS SOSS Order 2, 2.420–4.025 μm for NIRCam F322W2, 2.725–3.716 μm for NIRSpec G395H NRS1, 3.829–5.172 μm for NIRSpec G395H NRS2, and 0.518–5.348 μm for NIRSpec PRISM. Additionally, for NIRISS SOSS we exclude ~100 columns that are impacted by zeroth-order contamination from background sources[5]. The effect of resolution on the measured spectrophotometric transit depths is explored further below.

**Spectrophotometric Light Curve Fitting**
Across all instrumental modes, we fit the spectrophotometric light curves using the `Eureka!`[28] package, which jointly fits both a systematic and astrophysical model component to each of the light curves. The systematic model consists of a first-order polynomial in time, whereas the astrophysical transit models are computed using the `batman`[47] package. We also fit a step-function to the NIRSpec G395H data to account for the flux drop close to mid-transit in the uncorrected light curves that is driven by a mirror tilt event[12]. Orbital parameters are fixed during the fitting process using the values obtained from the white light curve fitting as shown in Table 1. Limb-darkening is incorporated using a quadratic law, and limb-darkening parameters are fixed in each of the light curve fits with initial values taken from the `ExoTIC-LD` package[48] using stellar parameters of [M/H]=0.0, $T_{\mathrm{eff}}$=5512, and $\log(g)$=4.7. Fitting for the limb-darkening parameters can produce wavelength-dependent biases as a function of wavelength binning resolution, and is investigated further below. The fitting itself is performed using MCMC as implemented by the `emcee`[49] package using 200 walkers, 1100 steps, and discarding the first 100 steps as a burn-in. Convergence is checked to ensure that the chains run for at least 50x the autocorrelation time.

The transit depth precision, native spectral resolving power, and wavelength coverage resulting from the spectroscopic light curve fitting are displayed alongside similar properties for archival *HST*, *VLT*, and *Spitzer* data[16,17] for WASP-39b in Extended Data Figure 3. Of the used *JWST* observational modes, NIRSpec PRISM provides the best transit-depth precision at all wavelengths, at the expense of greatly reduced resolution. Where the wavelength ranges of the higher resolution modes overlap, the NIRCam F322W2 provides the best precision from ~2.4–2.9 μm at a slightly lower resolution, as well as providing unique access from ~3.7–3.8 μm where the NIRSpec G395H has no sensitivity due to the gap between the detectors that the spectrum falls across. The NIRISS SOSS and NIRSpec G395H modes have similar resolutions in the narrow region where they overlap, with the NIRISS SOSS providing superior precision below ~2.75 μm. NIRSpec G395H has a similar precision to NIRCam F332W2 from ~3.0–3.5 μm despite having a factor of ~2 higher resolution due to it also having a factor of ~2 higher throughput.

It is clear that *JWST* provides a dramatic improvement on previous capabilities for the characterization of transiting exoplanet atmospheres, offering increased wavelength coverage, resolution and precision. The NIRISS SOSS observations provide superior resolution to existing *HST* infrared data at ~1–3 times higher transit-depth precision, and the NIRSpec PRISM offers a similar resolution at up to ~8 times higher precision. At longer wavelengths, all four instruments provide unrivaled advantages, and the *Spitzer* photometry is superseded by the spectroscopic capabilities of NIRSpec PRISM, NIRCam F332W2, and NIRSpec G395H. At native spectral resolution, the transit-depth precision of NIRCam F322W2 and NIRSpec G395H are a factor of ~2–3 times lower than *Spitzer*, but offer over two magnitudes improved resolving power compared to the *Spitzer* bandpasses. Furthermore, NIRSpec PRISM, in addition to the $R=100$ NIRCam F332W2 and NIRSpec G395H datasets, offer both ~2–3 times higher transit-depth precision and ~20–40 times improved resolving power. Nevertheless, ground-based telescopes and *HST* remain uniquely capable of accessing shorter wavelengths <0.5–0.6 μm, a wavelength range that is crucial for capturing and measuring the presence and strength of aerosol scattering and metal absorption lines[50].

**Wavelength Binning Investigation and a Dependence on Limb-Darkening**
We also investigate and extract transit spectra at coarser wavelength binning schemes following the procedure above at two to five times lower than native spectral resolution for all modes, as well as a $R=100$ binning scheme for the NIRISS SOSS, NIRCam F322W2, and NIRSpec G395H modes. We then bin the original native spectral resolution transmission spectra to an approximately similar resolution as each lower resolution spectrum to explore the prevalence and extent of resolution-dependent offsets. This is performed by taking the weighted mean of the native spectral resolution transit depths within each wavelength bin of the lower resolution spectrum. As each bin of the native spectral resolution spectrum constitutes multiple pixel columns, the wavelength ranges of the binned native resolution transit spectrum can differ slightly from those binned to a lower resolution prior to the light curve fitting. However, the focus of this investigation is to identify broad deviations between different resolutions, and a more detailed examination will require future analysis at the native pixel resolution. Residuals from each comparison, both when fitting for or fixing the quadratic limb-darkening parameters, are shown in Extended Data Figure 4.

In the case where the limb-darkening parameters are free parameters in the fitting process, we see significant differences between the binned native resolution spectra and those that are binned prior to the light curve fitting. Both NIRISS SOSS and NIRSpec G395H exhibit broad wavelength-dependent offsets that become more pronounced towards lower resolutions. Specifically, at R=100 NIRISS SOSS has a mean difference of 32±14 ppm (123±58 ppm above 2.2 μm) and NIRSpec G395H has a mean difference of 110±29 ppm (181±51 ppm above 4.5 μm). The NIRCam F332W2 exhibits a broad uniform offset at all wavelengths, with a mean difference of 58±19 ppm at R=100. Conversely, NIRSpec PRISM exhibits a non-significant mean difference of 3±14 ppm at 1/5 of the native spectral resolution. We re-emphasise that the wavelength ranges of the binned native resolution transit spectrum can differ slightly from those binned to a lower resolution prior to the light curve fitting. This is a likely driver of any observed narrow offset features, which are not explored in this work.

In the case where the limb-darkening parameters are fixed during the fitting process, the agreements between the different resolution spectra are drastically improved. Non-significant mean transit depth differences are exhibited by NIRISS SOSS (3±13 ppm, 8±53 ppm above 2.2 μm), NIRCam F332W2 (13±17 ppm), and NIRSpec PRISM (-3±12 ppm). In contrast, a wavelength-dependent offset does remain for NIRSpec G395H, although its mean offset of 37±27 ppm (87±48 ppm above 4.5 μm) is still reduced compared to when fitting for the limb-darkening parameters.

This stark difference in behavior when fitting or fixing the limb-darkening parameters is indicative of underlying biases of the adopted limb-darkening model. These biases appear to be strongest at regions of lower received detector counts, as evidenced by the offsets at the ends of the NIRSpec G395H / NIRISS SOSS data compared to the NIRCam F332W2 data (which has a relatively flat throughput and lower detector counts compared to other modes). Furthermore, these biases only seem to be evident for the higher resolution modes, and not the lower resolution NIRSpec PRISM mode. As the native spectral resolution light curves are at a lower signal-to-noise than those that were binned to lower resolutions prior to light curve fitting, it is likely that they are more susceptible to biases introduced when fitting for the limb-darkening parameters. As these biases are not necessarily Gaussian in nature (i.e. may not be comparable to additional random noise), consistent results are not seen between the binned native resolution spectra and those that are binned prior to the light curve fitting.

For future model analysis of these data, we recommend using the native spectral resolution spectrum for NIRSpec PRISM, the lower resolution R=100 spectra for the higher resolution modes, and in all cases the spectra that had their limb-darkening parameters fixed during the fitting process. We emphasize that this is not a global recommendation for all *JWST* datasets, but one that is specific to the current best understanding of these data. Fixing the limb-darkening parameters has provided greater agreement across different resolution binning schemes, however, the underlying reality is that these datasets are now uniformly biased by our limb-darkening assumptions. The extent of such biases is difficult to estimate at present, and significant future work will be required to explore the impact of different limb-darkening approaches on these data and those from other *JWST* observations.

**Wavelength Overlap Comparison Between Instruments**

Each of the four instrumental modes has an overlap between its wavelength coverage and the wavelength coverage of the other three modes, allowing for a comparison between their relative measurements of the transit depth. The broadest comparison comes from the NIRSpec PRISM mode, of which the wavelength coverage completely encompasses the coverage of the other modes, but at significantly lower resolution. As already shown in Figure 2, the NIRSpec PRISM data exhibit both a wavelength-independent offset across all wavelengths and a wavelength-dependent offset for data that experience saturation. Upon application of a -177 ppm uniform offset, as determined from the model analysis, the mean offset of NIRSpec PRISM is -124±6 ppm relative to NIRISS SOSS, 132±13 ppm relative to NIRCam F322W2, and 17±11 ppm relative to NIRSpec G395H. Although the NIRISS SOSS is in better agreement prior to offsetting the NIRSpec PRISM spectra, its wavelength range overlaps heavily with the saturated region of NIRSpec PRISM, which is not completely reliable. When looking at wavelengths unaffected by saturation, the mean offset of NIRSpec PRISM is -10±27 ppm relative to NIRISS SOSS. There are more significant discrepancies at localized regions of the wavelength coverage. In some cases—for example, the deviation at 2.6 μm—the difference can be attributed to the lower resolution of PRISM acting to "smooth over" atmospheric features that can be better captured at higher resolution.

Equivalent comparisons to Figure 2c for the smaller wavelength overlaps between the higher resolution NIRISS SOSS, NIRCam F322W2, and NIRSpec G395H modes are displayed in Extended Data Figure 5. We find an excellent agreement between NIRISS SOSS and NIRCam F322W2, with a non-significant mean offset of 11±49 ppm, compared to 32±46 ppm for the initial data release spectra. An offset is still present between NIRISS SOSS and NIRSpec G395H of -372±170 ppm, compared to -482±132 ppm for the initial data release spectra. However, this wavelength range is at the edge of the NIRSpec G395H and may be more significantly affected by systematic effects due to low throughput. An offset is also present between NIRCam F322W2 and NIRSpec G395H of -138±16 ppm, but this is greatly diminished compared to -343±16 ppm for the initial data release spectra. Furthermore, the distribution of residuals for NIRCam F322W2 versus NIRSpec G395H is close to the expected normal distribution, but with a uniform offset (Extended Data Figure 5). Given the agreement between NIRISS SOSS and NIRCam F322W2, this likely suggests that a wavelength-independent bias remains in the NIRSpec G395H spectrum even after performing a joint white light curve analysis.

All offsets as measured are presented in Extended Data Table 2 for ease of comparison. However, these values should not be interpreted as a generalizable property of the different detectors between the different instrumental modes. We predict that these offsets will be dependent on a currently unpredictable number of variables and will likely change between a given planet and observation. Instead, we emphasize that although offsets between JWST *spectra* have been identified, they can be mitigated through uniform light curve analysis.

In totality, these comparisons demonstrate that the joint white light curve analysis has dramatically improved the agreement between these various *JWST* datasets. Nevertheless, this improvement is not perfect, and some offsets do remain between datasets. Notably, these offsets are only present relative to either the NIRSpec PRISM or NIRSpec G395H modes. This may be an early indication of an uncorrected systematic or bias specific to the NIRSpec instrument, especially considering the excellent agreement between NIRISS SOSS and NIRCam F322W2. However, it is also possible that this agreement is a coincidence, and a firmer conclusion will require similar analyses across a wider range of *JWST* datasets. Where offsets are still present and saturation is not present, they appear to be close to normally distributed. Until an investigation even more detailed than that presented in this work is completed, the application of uniform offsets during model fitting and interpretation may be necessary.

**NIRSpec PRISM Saturation**

With a 2MASS *J* magnitude of 10.66[14], WASP-39 is above the brightness limit of the NIRSpec PRISM mode and produces detector saturation in the brightest pixel of the columns corresponding to ~0.63–2.06 μm. As the *JWST* detectors make use of non-destructive measurements to estimate the received flux (up-the-ramp sampling), if saturation occurs in a pixel after a large number of groups have been measured then the flux of that pixel may still be reliably measured with a sufficient number of unsaturated groups. However, for these NIRSpec PRISM observations there are only 5 groups per integration, with saturation occurring as early as the second group at the brightest part of the spectra. With so few measurements in each of these ramps, the ability to fit a slope and accurately estimate the flux for these pixels is diminished. When few groups are available, the linearity of the ramps is crucial to ensure an accurate determination of the count rates.

Extended Data Figure 6 demonstrates the differences between counts in neighboring groups as a diagnostic of the true linearity of the ramps. We see that the regions of the detector with higher count rates demonstrate a pattern that is indicative of an unexpectedly low count rate in the first group (2-1 is higher than 3-2, which suggests a similar effect to the first-group effect seen in Schlawin et al. 2023), or potentially an insufficient non-linearity correction (later group differences are lower than earlier group differences). Conversely, we see that rows 14 and 16 demonstrate a pattern that is indicative of a high count rate in the first group (2-1 is lower than 3-2), or potentially due to pixel cross-talk or charge diffusion as the central row approaches saturation in later groups (higher count differences in later groups than earlier groups). While we applied a more stringent saturation threshold in our updated PRISM analysis based on our analysis of the shapes of the ramps to avoid uncorrected non-linearity (approximately 70-75% rather than the 80% full-well threshold used in[11]), future work should more closely explore the accuracy of the NIRSpec non-linearity correction with data that saturate more slowly and therefore contain significantly more groups to better determine the shape of the ramp and first group impacts on partial saturation corrections. In particular, it would be useful to have multiple data sets of stars that saturate at different rates, in combination with a NIRISS SOSS observation, to fully characterize the impact of non-linearity, pixel cross-talk, and first group effects, on partially saturated exoplanet transit data. Such an analysis will have important implications on the full-well threshold that is appropriate to use when attempting to recover a partially saturated region on the NIRSpec detector and can help determine how and when cross talk occurs between neighboring pixels. We do note that regions of the detector that do not approach saturation in our data display flat group differences, which provides confidence in the extracted count rates in this region of the data.

The analysis of the differences of neighboring groups makes it clear that the measured stellar flux rate is likely not representative of the true stellar flux rate in regions of the detector that rapidly approach saturation. The combination of the two regimes in Extended Data Figure 6 together shape the measured count rates in the saturated region of the spectrum. As more groups are added, the impact of inaccurate group counts becomes less significant; however, for the saturated region, we are limited in the number of groups available. In particular, where a low first group dominates and few groups are available, the measured ramps are steeper than reality, corresponding to a higher extracted flux and a diluted transit depth. As shown in Figure 2, this is exactly what is seen in the saturated region of the offset NIRSpec PRISM transmission spectrum as compared to the NIRISS SOSS spectrum. To correct this effect we perform an analysis of how adding groups impacts the extracted spectrum to estimate the amount of excess flux measured within the saturated region. These excess flux measurements correspond directly to a dilution correction we can apply to the NIRSpec PRISM saturated region.

To explore how adding groups impacts the extracted spectrum, we reduce the data using the same number of groups across the entire wavelength range while masking the area that becomes saturated in that number of groups. This corresponds to a total of 5 median stellar spectra using 1, 2, 3, 4, or 5 groups. Extended Data Figure 7 shows these spectra relative to the 5-group spectrum, which is equivalent to our standard extraction. We see a trend of increasing extracted flux within the saturated region (vertical shaded gray regions, where the darkest region saturates after 1 group, and the lightest region saturates after 4 groups) when using a fewer number of groups, suggesting that the first group effect is biasing the ramps when less than 5 groups are used. Extrapolating this trend by fitting a Gaussian, we can determine the median excess flux in regions of the spectrum that saturate in 2, 3, or 4 groups. We do not report a correction for the region that saturates in 1 group because the 1-group spectrum is very noisy and the first-group effect is not well-enough understood. For this reason, we suggest avoiding saturation after only 1 good group.

Extended Data Figure 8 shows the residuals after subtracting the NIRISS SOSS spectra from the NIRSpec PRISM spectra before and after applying the dilution corrections shown in Extended Data Figure 7c. We find that the median differences between SOSS and PRISM reduce from $0.4\sigma$ to $0.08\sigma$. Though this is a marked improvement in agreement, it is important to note that the dilution corrections both assume that the shapes of the group differences are solely due to first-group effects and are extrapolated from a small number of groups, therefore suggesting that they may not be completely representative of the true effect. For this reason, and based on the extrapolated excess flux measurements, we suggest adopting a best practice of at least 5 groups before saturation is a safe regime in order to not be dominated by any of the effects that are impacting the group differences, and to ensure an accurate measurement of the stellar flux in this region.

We further caution on relying on applying a similar correction to other NIRSpec PRISM data on a similarly bright star without NIRISS SOSS data to compare to, particularly if the wavelength regime that saturates in your NIRSpec PRISM data is crucial to your science. While the broad wavelength coverage of NIRSpec PRISM is unmatched, the results in this work demonstrate that (1) it can be offset from other modes, although in this work it may be partially due to the low total number of groups, and (2) the feature sizes in the saturated region are unreliable. Even with the application of a dilution correction, larger spectral differences are present relative to the NIRISS SOSS data than are observed for all other modes relative to NIRISS SOSS. Our recommendations are to avoid partial saturation of NIRSpec PRISM, particularly if the saturated wavelengths are important to your science case, unless future *JWST* calibration data better understands the first-group effect and/or improves the non-linearity correction, therefore improving the ability to recover the saturated region. While the strategy of using multiple higher resolution modes would require the observation of a second transit, the higher spectral resolution may allow for additional science, such as the enhanced stellar modeling in Moran et al. 2023. NIRSpec PRISM remains a powerful tool for dimmer host stars to obtain wide wavelength coverage in an efficient single transit.

**Modeling**
To confirm the origin of the spectroscopic features present in the data from this synthesized release, we compare the observations against an atmospheric model of WASP-39b.
We utilize all of the spectroscopic *JWST* data presented in this work, except for data in the saturated region of NIRSpec PRISM. For NIRISS SOSS, NIRCam F322W2, and NIRSpec G395H, we use the data binned to $R$=100, and for NIRSpec PRISM we use the data at the native spectral resolution.

Motivated by the atmospheric inferences from the initial data releases (NIRISS: 10-30×solar metallicity, sub-solar C/O[5]; NIRSpec G395H: 3-10×solar metallicity, sub-solar C/O[12]; NIRCam: 1-100× solar metallicity, sub-solar C/O[7]; NIRSpec PRISM: ~10× solar metallicity, sub-solar C/O[11]) we choose a 10× solar metallicity, sub-solar carbon-to-oxygen ratio of 0.35 atmospheric composition under the assumption of full day-night heat redistribution. The atmospheric model assumes a 1-dimensional atmosphere under radiative-convective-photochemical equilibrium (1D-RCPE). Calculating the RCPE models corresponds to coupling a thermochemical solver with a kinetics solver as recently described in Bell et al. submitted. First, the model is computed using the ScCHIMERA radiative-convective equilibrium solver (RCE) introduced in Piskorz et al. 2018, with recent updates and implementations to *JWST* data from the initial data releases[5,11]. Then, the photochemical-equilibrium (PE) calculation corresponding to the atmospheric chemical state arising from the chemical kinetics due to photochemistry and vertical mixing is computed using the VULCAN[51] tool, following the model description in Tsai et al. 2022. We iterate over the RCE and PE calculations to ensure that the temperature-pressure profile and gas mixing ratios do not change, resulting in a ScCHIMERA-to-VULCAN-to-ScCHIMERA-to-VULCAN-to-ScCHIMERA computation chain.

We then considered the presence of inhomogeneous clouds and hazes by fitting a power-law and gray cloud-deck parametric model to the observations assuming the resulting 1D-RCPE model. The parametric cloud/haze model fits for a vertically uniform gray cloud opacity, $\kappa_{cld}$, and a power-law haze assuming a scaling law for its cross-section fitting for scattering slope, $\gamma$, and the scale, $a$[52]. Then, we allow for the presence of inhomogeneous cloud cover by using a linear combination of the cloudy/hazy model and the cloud-free model following the formalism in Line et al. 2016, with a cloud fraction $\phi$. When fitting the cloud parameters to the 1D-RCPE model, we allow for an offset in transit depth for the NIRSpec PRISM observations relative to all other instruments with a uniform prior between $\pm500$ ppm. Additionally, we allow for a scaling to the planetary radius referenced to 1 bar pressure. The final transmission spectrum in Figure 3 corresponds to the post-processed 1D-RCPE model with the median cloud parameters of $\log(\kappa_{cld})= -29.45$, $\gamma = 1.63$, $\log(a) = 1.95$, $\phi = 0.84$, and a scaling of the planetary radius of 97%.

The fit for cloud and haze properties suggests that a non-negligible offset must be applied to the NIRSpec PRISM observations to match the 1D-RCPE model. For this specific atmospheric composition, we find that a best-fit negative offset of 177 ppm is required to bring the data to the same transit depth level as the model. Comparing this model to the remaining data from other instruments (e.g., NIRISS SOSS O2) seems to suggest that additional offsets for each instrument and detector may be required. The presence of offsets in the data and their impact on the inferred atmospheric properties are explored in greater detail in the companion work of Welbanks et al.

**Data Availability**
The data used in this paper are associated with *JWST* program ERS 1366 (observations 1-4) and are available from the Mikulski Archive for Space Telescopes (https://mast.stsci.edu). Specific data products for the: time series spectra, white and spectroscopic light curve fits, transmission spectra, and model spectrum, can be accessed via Zenodo (DOI: 10.5281/zenodo.10161743).

**Code Availability**
This publication made use of the following code software to analyze these data: `NumPy`[53], `matplotlib`[54], `SciPy`[55], `pandas`[56,57], `Batman`[47], `emcee`[49], `Exotic-LD`[48], `dynesty`[45,58], `SpectRes`[59], `juliet`[31], and `Eureka!`[28].

**Acknowledgements**
This work is based on observations made with the NASA/ESA/CSA *JWST*. The data were obtained from the Mikulski Archive for Space Telescopes at the Space Telescope Science Institute, which is operated by the Association of Universities for Research in Astronomy, Inc., under NASA contract NAS 5-03127 for *JWST*. These observations are associated with program *JWST*-ERS-01366. Support for program *JWST*-ERS-01366 was provided by NASA through a grant from the Space Telescope Science Institute. This work is based in part on data collected under the NGTS project at the European Southern Observatory's La Silla Paranal Observatory. The NGTS facility is operated by a consortium of institutes with support from the UK Science

**Author contributions**

**Competing interests**

The authors declare no competing interests.

**Correspondence and requests for materials**

Should be addressed to A. L. Carter and/or E. M. May.

**Author Information**

*Corresponding authors' emails aacarter@stsci.edu, erin.may@jhuapl.edu

# Tables

**Table 1: Best-fit orbital and instrumental parameters from white light curve fitting.**

| Parameter | Value | Description |
|---|---|---|
| $P$ | $4.0552842 \pm^{0.0}_{0.0000035}$ | Orbital period [days] |
| $a/R_s$ | $11.390 \pm 0.012$ | Scaled semi-major axis |
| $b$ | $0.4498 \pm 0.0022$ | Impact parameter |
| $i$ | $87.7369 \pm 0.0024$ | Inclination [deg] |
| $e$ | 0 (fixed) | Eccentricity |
| $\rho_s$ | $1.6999 \pm^{0.0054}_{0.0055}$ | Stellar density [g cm$^{-3}$] |
| $T_{0,\text{NIRSpec/PRISM}}$ | $2459771.335647 \pm^{0.000013}_{0.000014}$ | NIRSpec PRISM mid-transit time* [days, BJD TBD] |
| $T_{0,\text{NIRCam}}$ | $2459783.5015000 \pm^{0.0000068}_{0.0000069}$ | NIRCam mid-transit time* [days, BJD TBD] |
| $T_{0,\text{NIRISS}}$ | $2459787.5567843 \pm^{0.0000073}_{0.0000073}$ | NIRISS mid-transit time* [days, BJD TBD] |
| $T_{0,\text{NIRSpec/G395H}}$ | $2459791.6120684 \pm^{0.0000094}_{0.0000089}$ | NIRSpec G395H mid-transit time* [days, BJD TBD] |
| $\sigma_{\text{NIRSpec/PRISM}}$ | $218.9 \pm^{7.0}_{6.8}$ | NIRSpec PRISM photometric jitter [ppm] |
| $\sigma_{\text{NIRCam/F322W2}}$ | $235.6 \pm^{10.9}_{10.2}$ | NIRCam F322W2 photometric jitter [ppm] |
| $\sigma_{\text{NIRCam/F210M}}$ | $75.5 \pm^{11.3}_{11.1}$ | NIRCam F210M photometric jitter [ppm] |
| $\sigma_{\text{NIRISS/Order 1}}$ | $110.3 \pm^{3.8}_{3.8}$ | NIRISS/SOSS Order 1 photometric jitter [ppm] |
| $\sigma_{\text{NIRISS/Order 2}}$ | $153.2 \pm^{8.1}_{7.9}$ | NIRISS/SOSS Order 2 photometric jitter [ppm] |
| $\sigma_{\text{NIRSpec/G395H/NRS1}}$ | $138.6 \pm^{6.5}_{6.0}$ | NIRSpec/G395H NRS1 photometric jitter [ppm] |
| $\sigma_{\text{NIRSpec/G395H/NRS2}}$ | $147.5 \pm^{7.6}_{7.0}$ | NIRSpec/G395H NRS2 photometric jitter [ppm] |

* A single time-of-transit was fitted for all datasets; we list the predicted time of transit for each instrument here based on that single fitted parameter.

# Extended Data Tables

## Extended Data Table 1: Overview of *JWST* Observations.

| Instrument | Observation Date | Wavelength Range (μm) | Resolving Power (λ/Δλ) | Number of Integrations | Groups per Integration | Integration Time (seconds) |
|---|---|---|---|---|---|---|
| NIRSpec PRISM | July 10th 2022 | 0.5 - 5.5 | ~100 | 21,500 | 5 | 1.13 |
| NIRCam F210M + F322W2 | July 22nd 2022 | 2.42 - 4.025 | ~1600 | 366 | 12 | 79.45 |
| NIRISS SOSS | July 26th 2022 | 0.6 - 2.8 | ~700 / ~1400 | 537 | 9 | 49.45 |
| NIRSpec G395H | July 30th 2022 | 2.725 - 3.716 3.829 - 5.172 | ~2,700 | 465 | 70 | 63.14 |

Observations are listed in the order they were acquired. For more detail on observational set up for these four modes, see[5,7,11]; and [12], respectively.

## Extended Data Table 2: Offsets Between Spectra.

| Instrument | NIRSpec PRISM | NIRcam F322W2 | NIRISS SOSS | NIRSpec G395H |
|---|---|---|---|---|
| NIRSpec PRISM | - | 132±13 ppm | -124±6 ppm | 17±11 ppm |
| NIRCam F322W2 | -132±13 ppm | - | -11±49 ppm | -138±16 ppm |
| NIRISS SOSS | 124±6 ppm | 11±49 ppm | - | -372±170 ppm |
| NIRSpec G395H | -17±11 ppm | 138±16 ppm | 372±170 ppm | - |

All offsets should be read as the selected spectrum in the left hand column relative to a spectrum of a different column. It is critical to emphasize that these offsets are unique to the spectra presented in this paper, and should not be interpreted as inherent offsets between JWST's instrumental modes.

# Figures

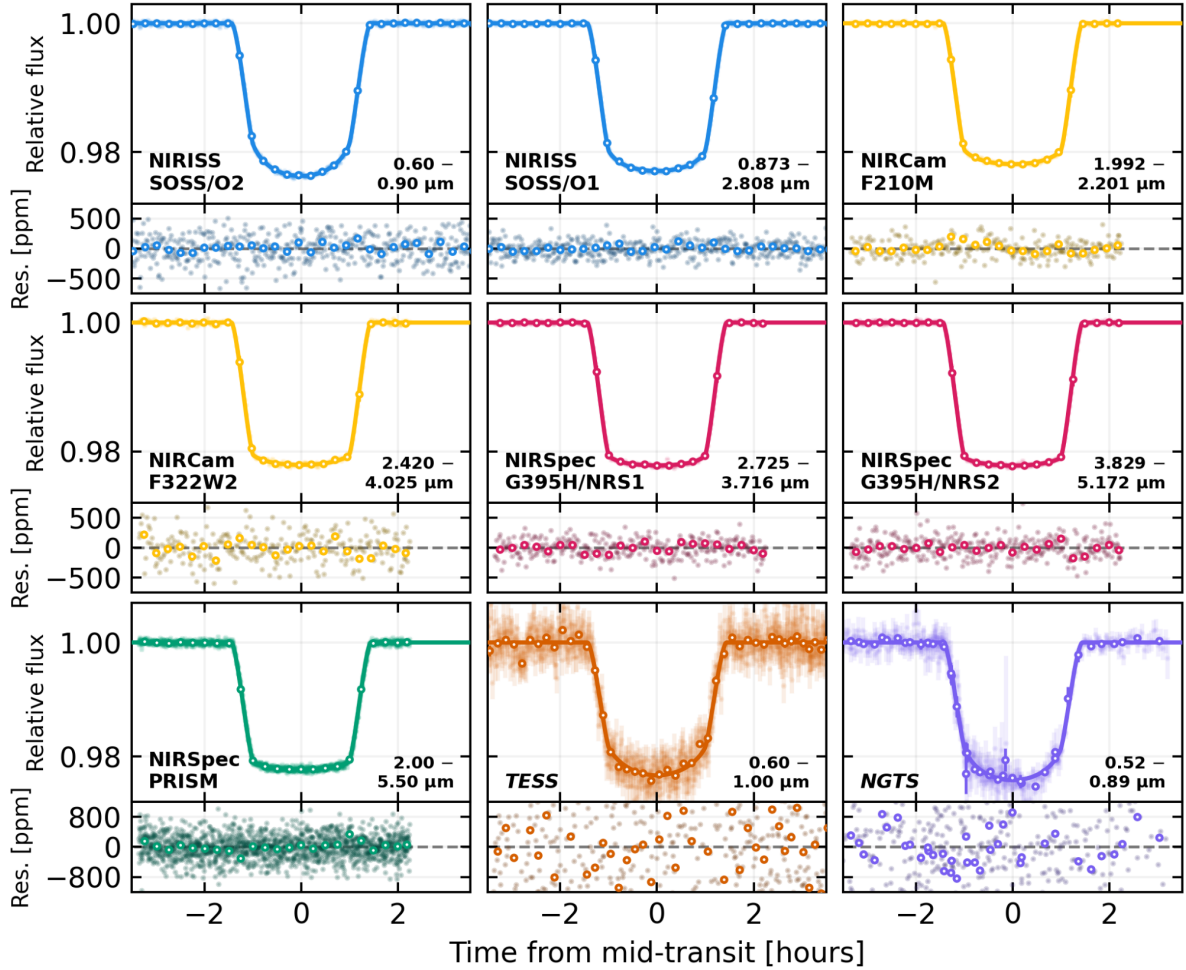

**Figure 1: White light curves of WASP-39b.** All data are presented after correcting for systematics, including the G395H mirror tilt event[12], with 1σ uncertainties. Solid lines indicate the best-fit model to each of the datasets from our joint fitting analysis, translucent points show the individual temporal measurements, and solid circles are the data after binning down to a lower temporal resolution (15 minutes for *JWST* datasets, 30 minutes for *TESS/NGTS*). The *JWST* light curve data are very precise and primarily lie underneath the best-fitting model lines. For the *TESS/NGTS* data, all light curves are phase-folded and the displayed best-fit model is an average across light curve fits. The residuals of the individual measurements compared to each best-fit model are displayed underneath each light curve.

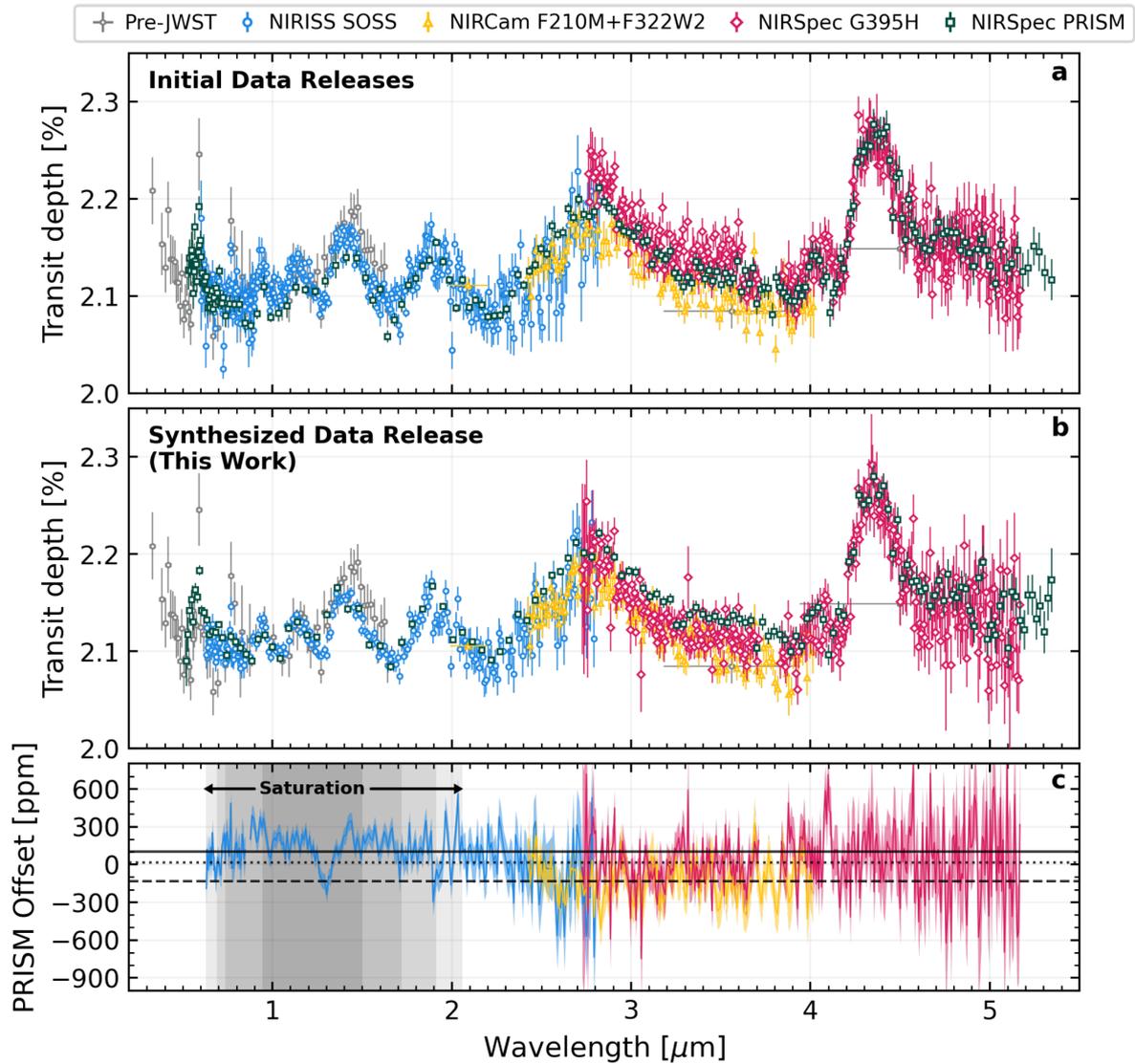

**Figure 2: The measured transmission spectrum of WASP-39b.** (**a**) The measured transmission spectra for all four instrumental modes as reported in the initial data release publications[5,7,11,12] (**b**) The measured transmission spectrum at native spectral resolution for NIRSpec PRISM, and at 1/5 of the native spectral resolution for the other modes (see Methods). (**c**) The residuals of the synthesized data for each mode relative to a linear interpolation of the NIRSpec PRISM data, with coloured shading indicating the 1σ uncertainty bounds. Horizontal lines indicate the median difference relative to the SOSS (solid), F322W2 (dashed), and G395H (dotted) data. See Methods for a detailed quantitative discussion of the offsets between individual modes. Regions where the NIRSpec PRISM data experience saturation are marked in gray shading, corresponding to saturation after one (darkest) to four (lightest) groups. All displayed uncertainties correspond to 1σ.

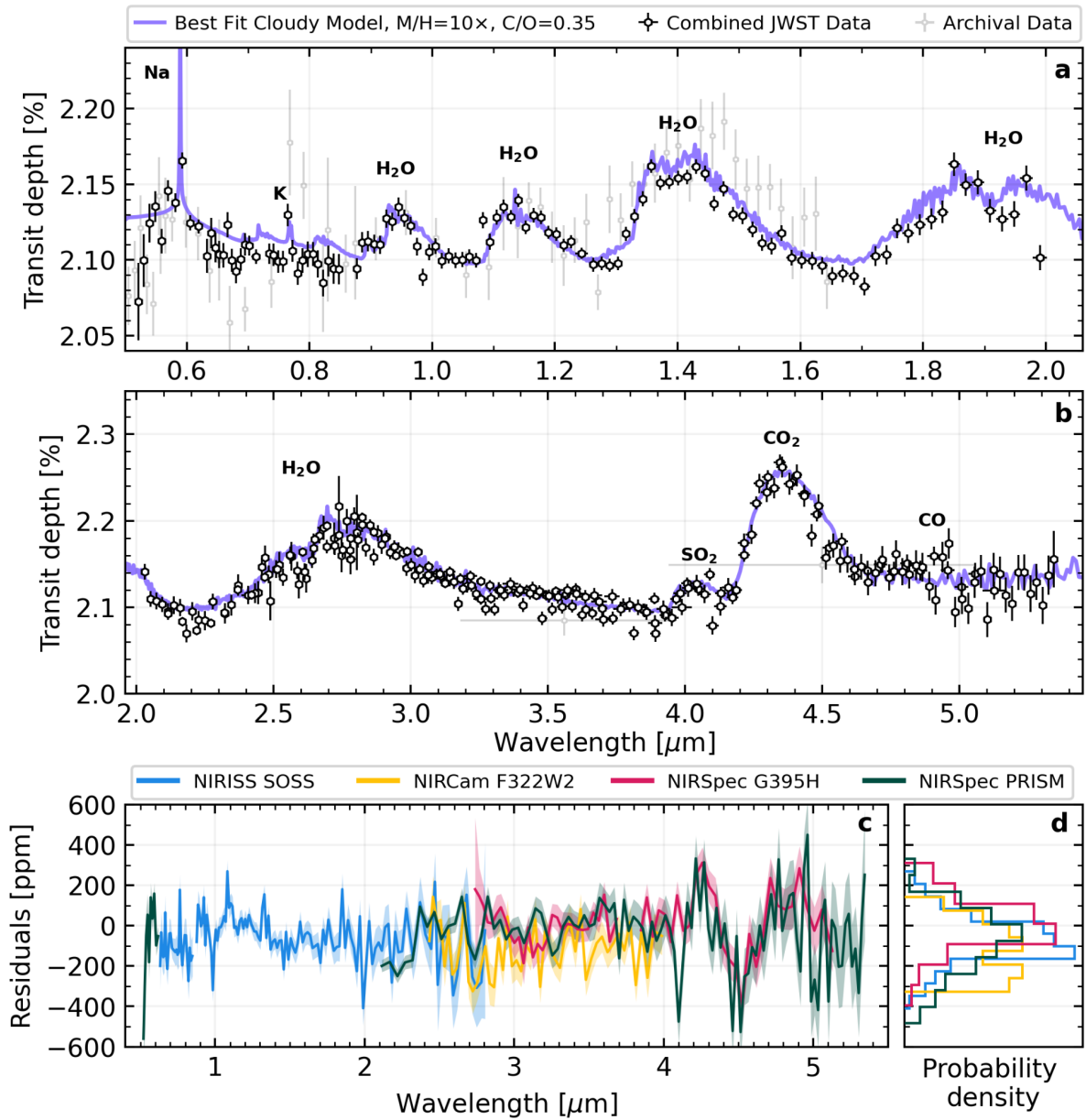

**Figure 3: Data comparison against a 1D RCPE model and residuals for the *JWST* spectrum of WASP-39b. (a,b)** The *JWST* transmission spectrum of WASP-39b with 1σ uncertainties (black hexagons), split at ~2.0 μm for clarity, alongside archival data (gray hexagons). For PRISM the data is at the native spectral resolution and excludes data in the saturated region; for the other modes the data is at $R$=100. The 10×solar metallicity, C/O=0.35 model is displayed in purple. **(c)** The residuals of the data for each mode relative to the best-fit model in addition to their 1σ uncertainties (shaded regions). **(d)** The probability density of the residuals for each mode relative to the best-fit model.

# Extended Data Figures

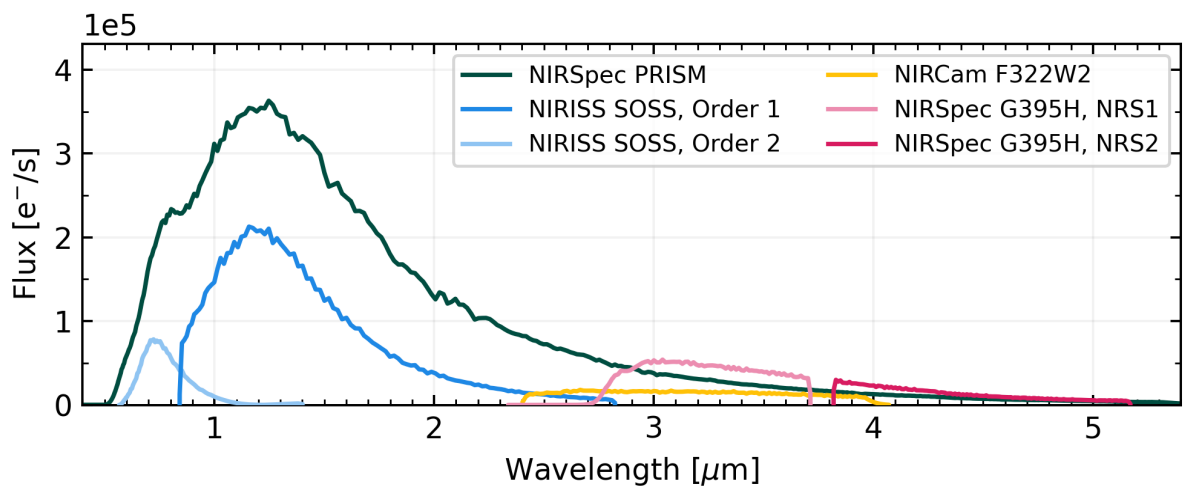

**Extended Data Figure 1: Median out-of-transit stellar spectra for WASP-39 in units of photoelectrons recorded at the detector.** All datasets have been binned in wavelength to match the NIRSpec PRISM data. Residual differences between datasets are a result of the different instrumental throughputs.

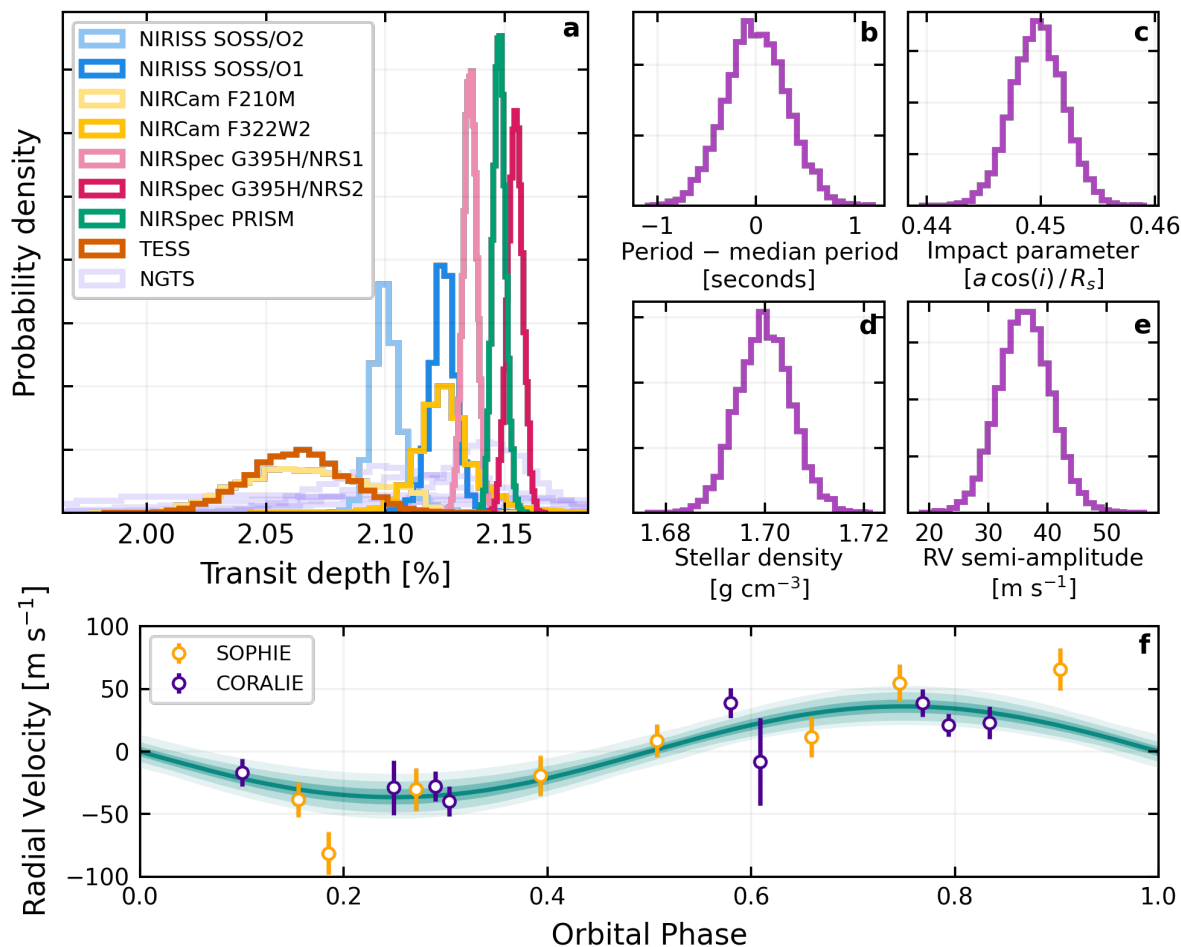

**Extended Data Figure 2: Best-fit orbital parameters for WASP-39b (a-e) and the radial velocity measurements (f).** The measured transit depth varies between instrumental modes as they each span a different wavelength range. The best-fitting model to the radial velocity data (solid line) alongside its 1, 2, and 3$\sigma$ contours (shaded regions) are also indicated.

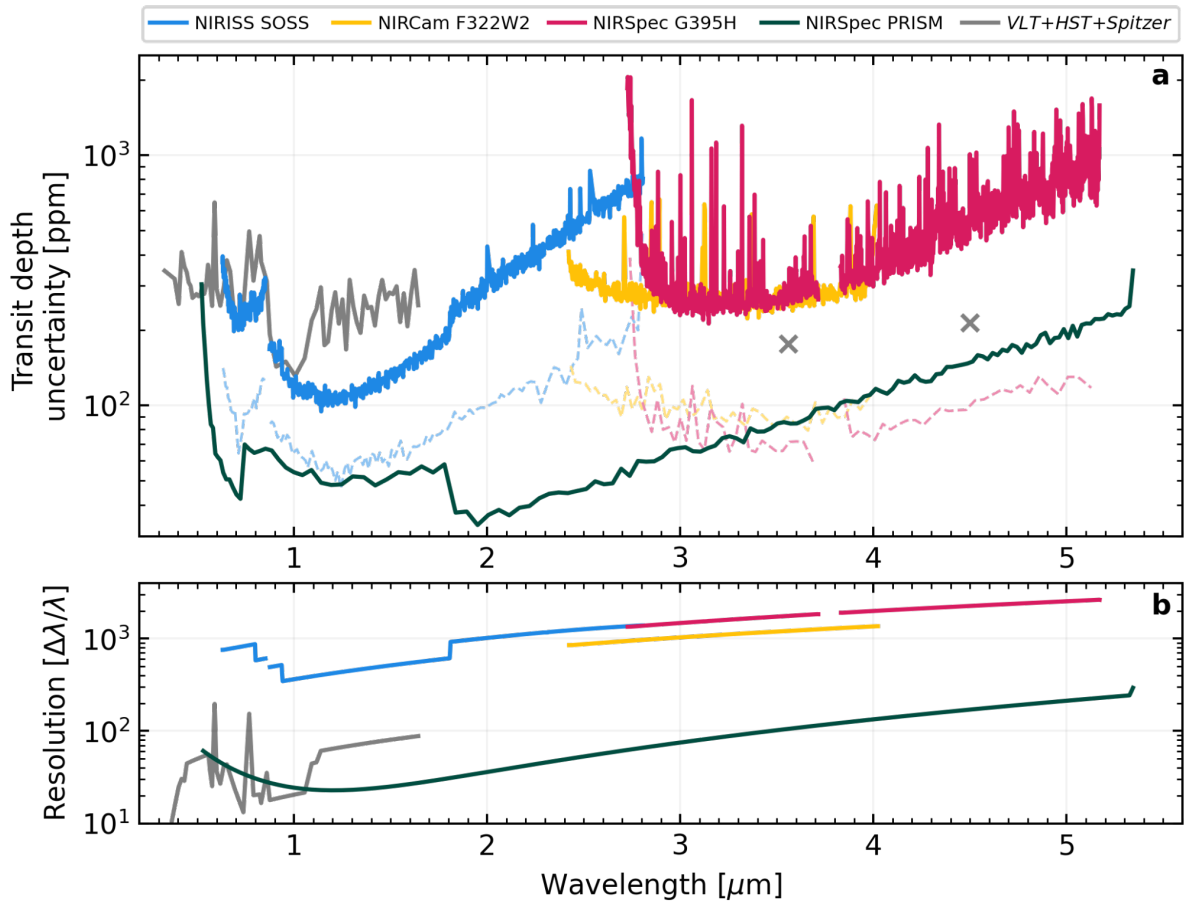

**Extended Data Figure 3: The achieved transit depth precision (a) and resolving power (b) across all datasets.** For the transit depth uncertainty, solid lines correspond to the native spectral resolution datasets, and dashed lines correspond to the *R*=100 datasets. Archival *HST*, *VLT,* and *Spitzer* data are also displayed (gray lines and crosses). As the reported archival transit depths are an averaged combination of multiple transits, we inflate the reported transit depth uncertainties by a factor of $\sqrt{N}$, where $N$ is the number of transits for a given bandpass, to more accurately compare the signal-to-noise provided by a single transit across different instruments.

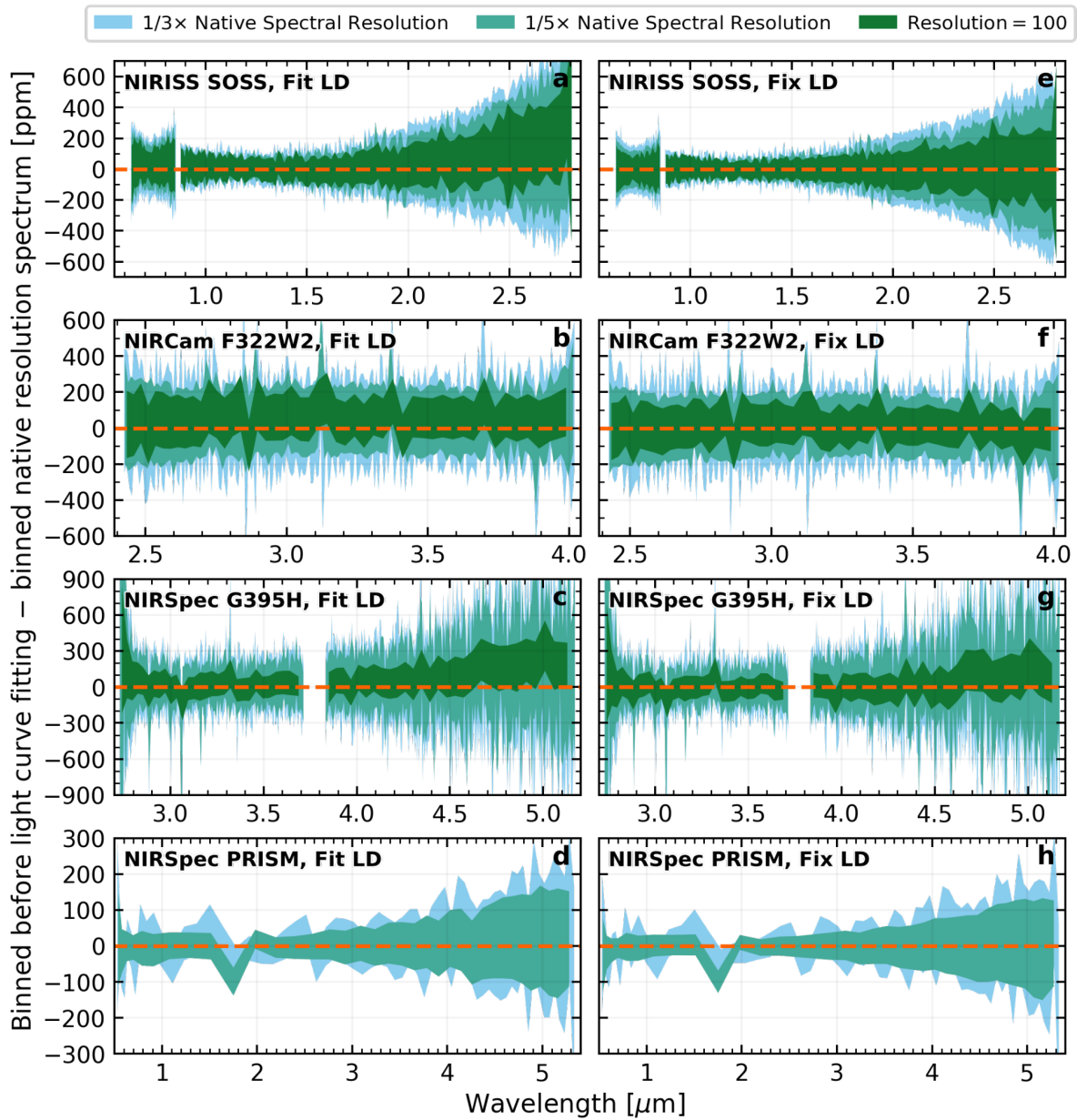

**Extended Data Figure 4: Comparison between the transmission spectra under different binning schemes and limb-darkening approaches for each instrumental mode.** Each panel shows the residual when comparing the spectroscopic transit depths as determined by binning before light curve fitting is performed versus those determined by directly binning the measured transmission spectrum at the native spectral resolution. The width of each shaded region corresponds to the 1$\sigma$ uncertainty in the measured residual. Dashed lines indicate zero deviation.

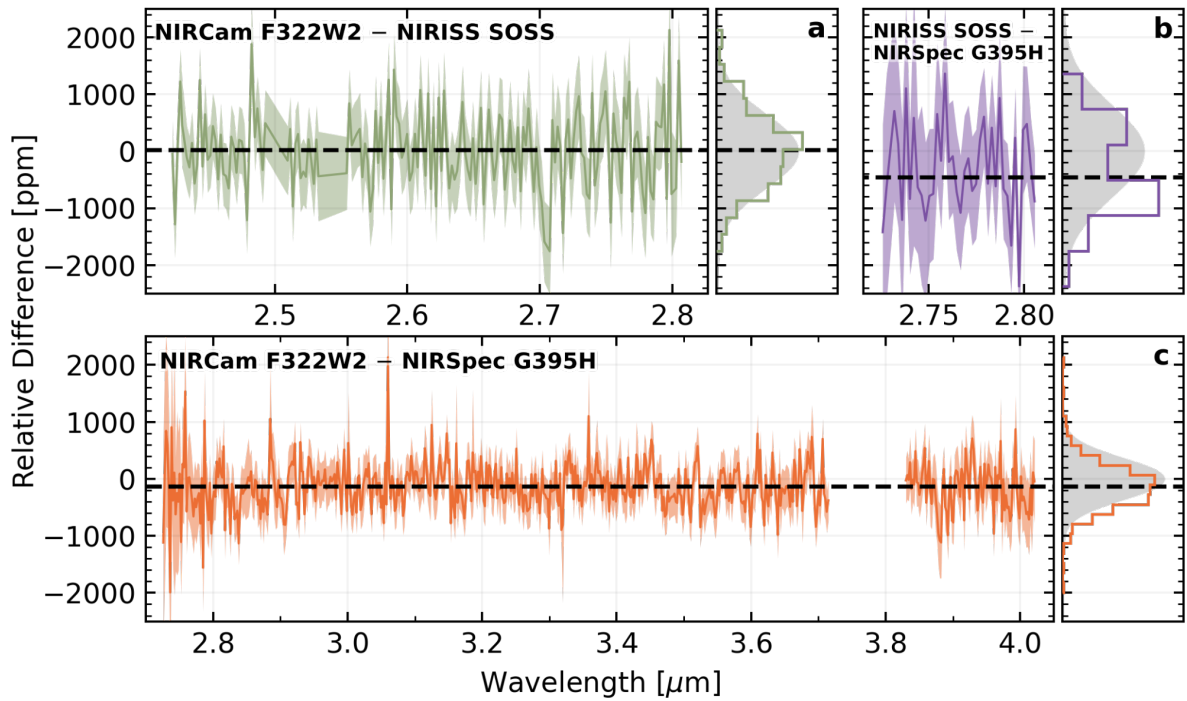

**Extended Data Figure 5: Comparison of *JWST* instrumental modes across their overlapping wavelength regions.** Due to differences in wavelength binning, the minuends are interpolated to the wavelength grid of the subtrahends. Solid lines indicate the residual between the measured transit depths of the indicated instrumental modes at their native resolution, and the dashed line is the median difference across all wavelengths. A similar comparison relative to NIRSpec PRISM mode is displayed in Figure 2c. The probability density of the residuals is displayed next to each comparison, alongside a Gaussian distribution with zero mean and a standard deviation equal to the median error on the residual (gray shading).

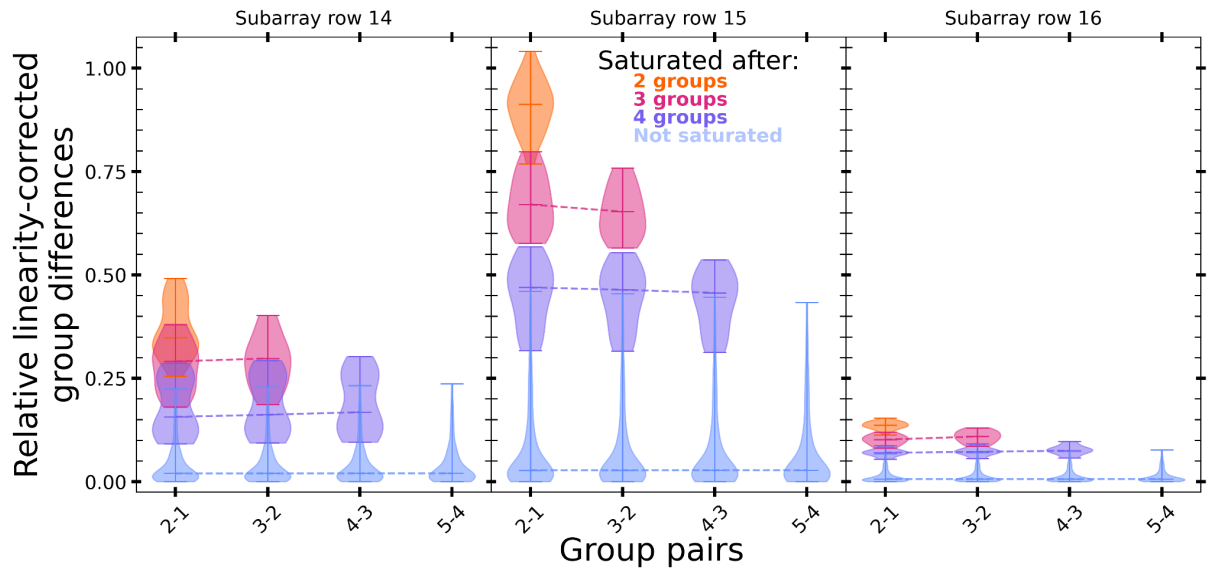

**Extended Data Figure 6: Differences in counts between neighboring groups after the non-linearity correction.** A linear ramp would show a flat line difference between neighboring groups. Residual slopes in differences between neighboring groups are indicative of detector effects that are not fully corrected (denoted by the dashed lines connecting the median of group pairs). Above we show the distributions in differences between neighboring groups for three detector rows where row 15 corresponds to the central location of the trace, and rows 14 and 16 are the neighboring detector rows. Distributions are colored by the number of groups that are usable before reaching our saturation threshold. Note that we follow the same methodology as[11] in flagging an entire column as saturated when the central row saturates.

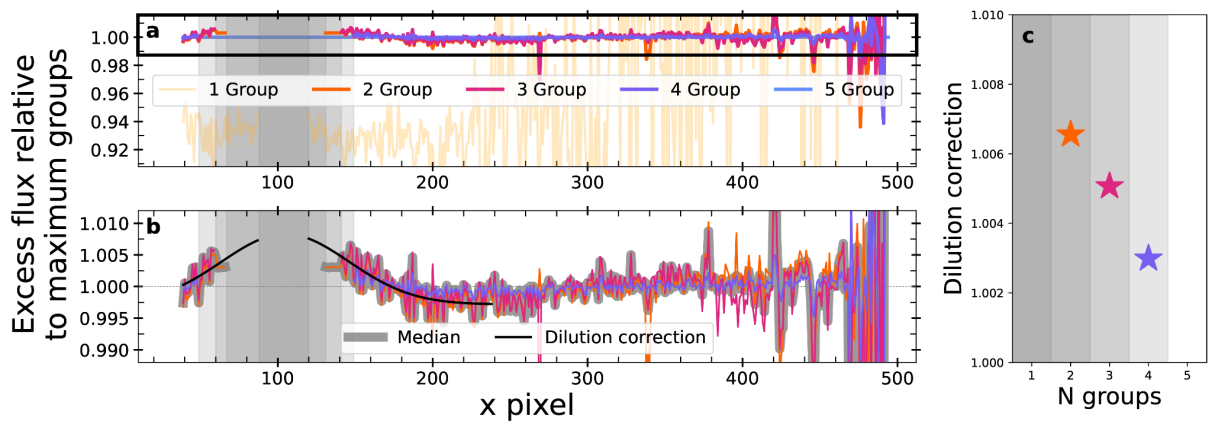

**Extended Data Figure 7: Analysis of excess flux relative to the nominal 5-group spectrum.** The 5-group spectrum uses a varying number of groups within the saturated region, as is done in the standard PRISM extraction. **(a)** Extracted spectra using different numbers of groups relative to the 5-group spectrum. Data are not shown in regions that are already saturated, or where the 5-group spectrum uses the same number of groups. **(b)** Zoomed-in view of the Top Left panel, specifically the region in panel a that is enclosed in a box, excluding the noisy 1-group spectrum. A Gaussian is fit to the excess flux spectrum. **(c)** The dilution corrections derived from the median excess flux fit. In all panels the vertical-gray shaded regions denote the level of saturation, with the darkest corresponding to saturation after 1 group and the lightest corresponding to saturation after 4 groups.

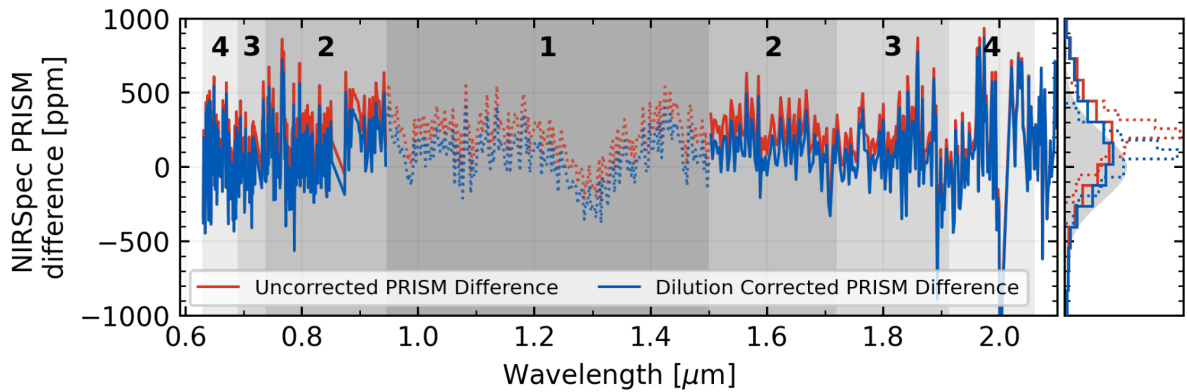

**Extended Data Figure 8: Comparison of NIRISS SOSS to NIRSpec PRISM before and after the estimated dilution correction.** Shaded regions indicate the wavelength ranges that experience saturation after the number of groups indicated in bold. A dilution correction cannot be computed for the region that experiences saturation after one group (dotted lines); however, we show an example correction using the correction value for the regions experiencing saturation after two groups. The probability density of the residuals is displayed in the right panel, alongside a Gaussian distribution with zero mean and a standard deviation equal to the median error on the residual (gray shading). We note that the downwards peak around 1.3 micron is likely due to this spectral region having the lowest resolution across the instrument, resulting in the highest counts in pixels at these wavelengths such that these detector columns get closer to saturation within the first group. Due to insufficient non-linearity correction within the *JWST* pipeline and the first-group effect, lower-than-expected flux is measured in this region, resulting in deeper transit events.